# Density, porosity and magnetic susceptibility of the Košice meteorite shower and homogeneity of its parent meteoroid

Tomáš Kohout[1,2*], Karol Havrila[3], Juraj Tóth[3], Marek Husárik[4], Maria Gritsevich[5,6,7,8], Daniel Britt[9], Jiří Borovička[10], Pavel Spurný[10], Antal Igaz[11], Ján Svoreň[4], Leonard Kornoš[3], Peter Vereš[3,12], Július Koza[4], Pavol Zigo[3], Štefan Gajdoš[3], Jozef Világi[3], David Čapek[10], Zuzana Krišandová[4], Dušan Tomko[4], Jiří Šilha[3], Eva Schunová[3], Marcela Bodnárová[4], Diana Búzová[13] and Tereza Krejčová[4,14]

1. *Department of Physics, University of Helsinki, P.O. Box 64, 00014 Helsinki University, Finland*
2. *Institute of Geology, Academy of Sciences of the Czech Republic, Rozvojová 269, 16500 Prague 6, Czech Republic*
3. *Faculty of Mathematics, Physics and Informatics, Comenius University, Mlynská dolina, SK-84248 Bratislava, Slovakia*
4. *Astronomical Institute, Slovak Academy of Sciences, SK-05960 Tatranská Lomnica, Slovakia*
5. *Finnish Geodetic Institute, Geodeetinrinne 2, P.O. Box 15, FI-02431 Masala, Finland*
6. *Ural Federal University, Mira st.19, 620002 Ekaterinburg, Russia*
7. *Dorodnicyn Computing Centre, Russian Academy of Sciences, Vavilova ul. 40, 119333 Moscow, Russia*
8. *Institute of Mechanics, Lomonosov Moscow State University, Michurinsky prt., 1, 119192, Moscow, Russia*
9. *Department of Physics, University of Central Florida, P.O. Box 162385-2385, Orlando, 32816-2385, FL, USA*
10. *Astronomical Institute, Academy of Sciences of the Czech Republic, CZ-25165 Ondřejov, Czech Republic*
11. *Hungarian Astronomical Association, MCSE, P.O. Box 219, H-1461 Budapest, Hungary*
12. *Institute for Astronomy, University of Hawaii at Manoa, HI 96814, USA*
13. *Department of Biophysics, Faculty of Science, P.J.Šafarik University, Košice, Slovakia*
14. *Department of Theoretical Physics and Astrophysics, Masaryk University, Kotlářská 2, 61137 Brno, Czech Republic*

\* Corresponding author e-mail: tomas.kohout@helsinki.fi; phone: +358919151008; fax: +358919151000

**Abstract**

Bulk and grain density, porosity, and magnetic susceptibility of 67 individuals of Košice H chondrite fall were measured. The mean bulk and grain densities were determined to be 3.43 g/cm$^3$ with standard deviation (s.d.) of 0.11 g/cm$^3$ and 3.79 g/cm$^3$ with s.d. 0.07 g/cm$^3$, respectively. Porosity is in the range from 4.2 to 16.1%. The logarithm of the apparent magnetic susceptibility (in 10$^{-9}$ m$^3$/kg) shows narrow distribution from 5.17 to 5.49 with mean value at 5.35 with s.d. 0.08.

These results indicate that all studied Košice meteorites are of the same composition down to ~g scale without presence of foreign (non-H) clasts and are similar to other H chondrites. Košice is thus a homogeneous meteorite fall derived from a homogeneous meteoroid.

**1 Introduction**

In this study, we tested homogeneity of the recovered Košice meteorites through measurement of their bulk and grain density, porosity, and magnetic susceptibility. The meteoroid entered the Earth's atmosphere on February 28, 2010 over Slovakia and caused an exceptionally bright fireball with corresponding sonic booms. Most of the Košice meteorite samples used in our analyses were recovered North-West of Košice city, Eastern Slovakia within 4 weeks of the actual fall. The meteorite fall and its recovery are described in detail by Borovička et al. (2013) and Tóth et al. (2014). Mineralogical analyses of the meteorite classifying it as H5 chondrite can be found in Ozdín et al. (2014). Based on the fragment distribution and fragmentation modelling Košice meteoroid (Gritsevich et al., 2014) may have been composed of two individual bodies.

The data presented in this manuscript are unique in the sense that the measurements were done systematically on a large number (67 samples) of freshly recovered individuals of the same shower, with known heliocentric orbit. The extent of our data set significantly exceeds the number of fragments which underwent conventional chemical and mineralogical analyses and allow us to test the homogeneity of the Košice meteoroid.

## 2 Material and methods

Density and porosity data were obtained using a mobile laboratory described in Kohout et al. (2008). Bulk volume was determined using a modified Archimedean method (Consolmagno and Britt, 1998; Macke et al., 2010) incorporating glass beads ~ 0.3 mm in diameter. Ten sets of measurements per sample were done and each sample was measured independently by at least two persons. The method was thoroughly tested and calibrated prior the measurements using volume standards and the resolution and precision was determined to be ± 0.1 cm$^3$. Grain volumes of the meteorites were measured using a Quantachrome Ultrapyc 1200e Helium pycnometer. The resolution and precision of this device is better than ±0.01 cm$^3$. The relative error of both volumetric methods increases with decreasing sample size. However, compared to bulk volume, the grain volume is determined with ten times higher resolution and precision. Masses were determined using a digital OHAUS Navigator and OHAUS Scout scales with 0.1 g and 0.01 g resolution and precision, respectively. Balances were calibrated prior the measurements using internal calibration procedure (OHAUS Navigator) or mass standards (OHAUS Scout).

Magnetic susceptibility of samples smaller than 2.5 cm was measured using a ZH instruments SM-100 susceptibility meter frequency range of 0.5-8 kHz and 10-320 A/m RMS field amplitude. A frequency of 1 kHz and field amplitude of 320 A/m were used for routine measurements. Two samples were tested for frequency and field amplitude dependence using the same instrument. The frequency and field ranges were cross-calibrated using a ferrite standard prior to the measurements. For larger samples, a Hämäläinen TH-1 portable susceptibility meter with a large 12 cm coil was used.

Susceptibility of the samples was measured three times along three perpendicular directions. Subsequently, a logarithm of the apparent magnetic susceptibility (in $10^{-9}$ m³/kg) was calculated as described in Rochette et al. (2003) and can be used as a proxy for a concentration of the metallic iron (Rochette et al., 2003). Relative error in the determined value of the magnetic susceptibility logarithm is below 3%. For samples with bulk volume information available a true susceptibility was also calculated using an ellipsoid shape correction after Osborn (1945).

For density, porosity, and susceptibility statistical calculations, only samples larger than 5 g were considered.

## 3 Results

### 3.1 Density and porosity of the Košice meteorites

The results of the density and porosity measurements of the Košice meteorites are summarized in Table 1. The bulk density of individual pieces ranges from 3.15 to 3.64 g/cm³ with the mean value of 3.43 g/cm³ and standard deviation (s.d.) of 0.11 g/cm³. This is in good agreement with bulk density of other H chondrites (both falls and finds average, 3.42 s.d. 0.18 g/cm³) reported in Consolmagno et al. (2008).

The grain density of the individual pieces ranges from 3.62 to 3.92 g/cm³ with the mean value of 3.79 s.d. 0.07 g/cm³. This is again in a very good agreement with bulk density of other H chondrites (falls average 3.72 s.d. 0.12 g/cm³) reported in Consolmagno et al. (2008).

The porosity of the individual pieces ranges from 4.2 to 16.1% with the mean value of 9.88 % s.d. 3.01% which is slightly higher compared to other H chondrites (falls average, 7.0% s.d. 4.90%) reported in Consolmagno et al. (2008), closer to the value of 9.4% s.d. 0.5% in updated study by Macke (2010).

Smaller samples show higher density and porosity scatter (Figs. 1 and 2). This is most likely a combination of higher relative uncertainty in the measured values of smaller samples as described above and increasing inhomogeneity of the material at smaller scales. In general, the larger samples have lower porosities compared to smaller ones. This can be explained by their origin through fragmentation of parent meteoroid upon atmospheric entry. Looser, more porous parts of the parent meteoroid tend to fragment progressively into more numerous smaller fragments, while more coherent, less porous parts of the parent meteoroid are more resistant to fragmentation, producing larger fragments. However, in order to get deeper insight into the mechanism behind porosity vs. meteorite size trend, future systematic research of other meteorite showers is needed.

## 3.2 Magnetic susceptibility of the Košice meteorites

It has been reported previously that various meteorite types can be distinguished from their magnetic susceptibility (e.g. Kukkonen and Pesonen, 1983; Terho et al. 1993a, 1993b; Rochette et al., 2003; Smith et al., 2006; Kohout et al., 2008) and thus can be used as a compositional homogeneity indicator of a meteorite shower (Consolmagno et al., 2006; Kohout et al., 2010). The results of the susceptibility measurements done on the Košice meteorites are summarized in Table 1. The logarithm of the apparent magnetic susceptibility (in $10^{-9}$ m$^3$/kg) shows a narrow distribution from 5.17 to 5.49 with the mean value of 5.35 s.d. 0.08. This is in close agreement with the other H chondrite falls (5.32 s.d. 0.10 average reported in Rochette et al. (2003) or 5.29 s.d. 0.10 average reported in Smith et al. (2006). The values are quite uniformly distributed among samples of various masses (Fig. 3). There is only a slight increasing trend observed in susceptibility with increasing mass. However, the low and high mass values are represented by only a few samples and thus reliability of this trend is questionable. Similarly, as with density values, smaller samples show slightly higher susceptibility scatter, most likely due to inhomogeneity in composition and iron distribution at smaller scales. No correlation was observed between magnetic susceptibility and grain density (Fig. 4).

The test for frequency and field dependence of magnetic susceptibility showed different results for the two tested samples (no. 1 and 2). In meteorite no. 1, there was almost 30% susceptibility decrease between 0.5 and 1 kHz measurement frequencies and 10% susceptibility increase between 10 and 320 A/m measurement fields. However, the same measurements resulted in susceptibility changes below 5% for meteorite no. 2. Thus, it seems the grain size of magnetic minerals is not uniform in these two samples. The enhanced frequency dependence in sample no.1 can be interpreted as enhancement in concentration of metal nanoparticles in superparamagnetic state. The field dependence can be interpreted as contribution of large multidomain grains.

## 4 Discussion

Due to our non-destructive measurement technique it was possible to determine the physical properties of 67 individual Košice meteorites. This gives us a unique opportunity to test the homogeneity of the recovered meteorite material and the Košice parent meteoroid. Because conventional mineralogical and chemical analyses are destructive, these data are available only from 8 individuals of the Košice shower (Ozdín et al., 2014). Observed similarity, particularly in grain density and magnetic susceptibility, among analyzed samples and other unanalyzed Košice individual pieces gives us confidence to conclude that the Košice shower and its parent meteoroid seems to be homogeneous down to ~ g scale without evidence of presence of non-H fraction. If the Košice meteoroid was composed of two individual bodies as suggested by Gritsevich et al. (2014), these two bodies were of identical composition and most likely separated in past either by impact processes, thermal stress, or rotational forces.

The Košice physical properties are a close match to the other H chondrites (Consolmagno et al., 2006, 2008) and H chondrite showers, for example Buzzard Coulee, Grimsby, or Pultusk. Two size-dependent trends can be observed from Fig. 1 and 2. Scatter in the bulk and grain density slightly increases with decreasing size. This can be explained by the coarse-grained (~ mm) fabric of the material. It may be also partly affected by fact that relative error in volume measurement is increasing with decreasing sample size as described in Materials and Methods section. Additionally, the larger samples tend to have lower porosity, most likely because they are derived from the more coherent parts of the parent meteoroid.

In general, H chondrite mineralogy (e.g. Urey and Craig, 1953), oxygen isotope chemistry (Clayton et al., 1991), formation age (Rb-Sr age of 4.56 Ga, Kaushal and Wetherill, 1969), and cosmic ray exposure ages (pronounced peak at 8 Ma, Marti and Graf, 1992) of H chondrites is relatively uniform suggesting single parent body for all H chondrites. Asteroid 6 Hebe has been suggested as a candidate of such body (Gaffey and Gilbert, 1998).

However, not all H chondrites arrive to the Earth in a homogeneous meteorite fall implying on extensive post-parent body processing. Meteorite polymict breccias including H chondrite material are summarized in Bischoff et al. (2006). Recently, H chondrites were reported, along with L, LL and E chondrites, to be mixed with the dominant ureilites in the Almahatta Sitta meteorite shower originating from the fall of asteroid 2008 $TC_3$ (Bischoff et al., 2010; Kohout et al., 2010; Zolensky et al., 2010) or part of the recently recovered Benešov meteorite consisting of both H and E chondrite clasts (Spurný et al., 2012). Additionally, the orbit of Příbram H5 meteorite parent meteoroid (Ceplecha, 1961) closely resembles that of Neuschwanstein EL6 meteorite parent meteoroid (Spurný et al., 2003) what may be caused by a stream of heterogeneous meteoroids directing various types of meteoritic material towards the Earth (Kornoš et al., 2008); possibly caused by tidal disruption of a near-Earth rubble pile asteroid in the close vicinity of Earth (Tóth et al., 2011). However, no evidence for such polymict breccia is observed in the Košice meteorite shower.

## 5 Conclusions

The individual samples of the Košice meteorite shower seem to be homogeneous down to ~ g scale. Their physical properties are similar to other H chondrites. Based on the uniform narrow distribution of grain density and magnetic susceptibility, we can conclude that all studied meteorites are of the same H chondrite composition and are similar to other H chondrites. No foreign (non-H) clasts were detected. This makes Košice a homogeneous fall derived from a homogeneous parent meteoroid. In the case of binary nature of Košice meteoroid, both components are indistinguishable in composition and were most likely mechanically broken apart in past from a single body.

The scatter in bulk and grain density is slightly increasing with decreasing size, most likely due to the coarse-grained (~ mm) fabric of the material. The larger samples tend to have lower porosity, most likely because they are derived from more coherent parts of the parent meteoroid.

## 6 Acknowledgements

We would like to thank Oskar Öflund foundation for travel support. The laboratory work was supported by Ministry of Education, Youth and Sports of the Czech Republic LH12079, VEGA 1/0636/09, 2/0022/10, and APVV-0516-10 grants, and Academy of Finland project 257487 and 260027. MG was supported by Emil Aaltonen foundation post-doc grant. TK was supported during his stay in AI SAS by the National scholarship program of Slovak Republic (SAIA). We would like also to thank the staff of the Astronomical Observatory of Comenius University at Modra, where measurements and analyses were done in July 2011.

Table 1. Summary of the Košice meteorite measurements. Individuals are sorted by increasing mass. Values in italics come from pieces smaller than 5 g and were not considered in statistics due to higher uncertainty in the values. $\rho_B$ – bulk density, $\rho_G$ – grain density, p – porosity, $\chi_{mA}$ – apparent mass susceptibility, $\chi_{mT}$ – true mass susceptibility, $\chi_{VT}$ – true volume susceptibility.

| Meteorite no. | Mass (g) | $\rho_B$ (g/cm³) | $\rho_G$ (g/cm³) | p (%) | $\chi_{mA}$ (10⁻⁹ m³/kg) | log $\chi_{mA}$ (in 10⁻⁹ m³/kg) | $\chi_{mT}$ (10⁻⁹ m³/kg) | $\chi_{VT}$ (10⁻³ SI) |
|---|---|---|---|---|---|---|---|---|
| *47* | *0.56* | | *3.7* | | *149213* | *5.17* | | |
| *69* | *0.66* | | *4.0* | | *191723* | *5.28* | | |
| *68* | *1.15* | | *3.8* | | *196343* | *5.29* | | |
| *51* | *1.78* | | *3.8* | *-* | *228000* | *5.36* | | |
| *34* | *1.86* | | *3.7* | | *164500* | *5.22* | | |
| *74* | *2.38* | | *4.3* | | | | | |
| *14* | *2.69* | | *3.8* | | *176357* | *5.25* | | |
| *3* | *2.72* | *3.9* | *4.1* | *6* | *185480* | *5.27* | *244970* | *943* |
| *45* | *2.93* | | *4.0* | | *203090* | *5.31* | | |
| *48* | *3.02* | | *3.9* | | *189500* | *5.28* | | |
| *37* | *3.18* | | *3.7* | | *201910* | *5.31* | | |
| *76* | *3.85* | | *3.8* | | *229500* | *5.36* | | |
| *9* | *3.90* | *3.2* | *3.9* | *20* | *183990* | *5.26* | *224869* | *711* |
| *59* | *3.93* | | *3.8* | | | | | |
| *35* | *3.95* | | *4.0* | | *171800* | *5.24* | | |
| *31* | *4.04* | | *3.9* | | *247590* | *5.39* | | |
| *8* | *4.55* | *3.0* | *3.8* | *20* | *198217* | *5.30* | *245638* | *747* |
| *29* | *4.69* | | *4.0* | | *186600* | *5.27* | | |
| 41 | 5.10 | 3.3 | 3.8 | 16 | 203477 | 5.31 | 258824 | 852 |
| 6 | 5.75 | 3.6 | 3.9 | 8 | 234300 | 5.37 | 325960 | 1157 |
| 11 | 6.00 | 3.2 | 3.8 | 14 | 181197 | 5.26 | 223132 | 718 |
| 42 | 6.01 | 3.3 | 3.8 | 12 | 267723 | 5.43 | 369264 | 1230 |
| 7 | 6.33 | 3.6 | 3.9 | 10 | 189543 | 5.28 | 236875 | 848 |
| 44 | 6.41 | 3.5 | 3.9 | 11 | 250257 | 5.40 | 347220 | 1198 |
| 30 | 6.46 | 3.2 | 3.7 | 15 | 148830 | 5.17 | 175499 | 553 |
| 62 | 6.55 | 3.6 | 3.8 | 4 | 265657 | 5.42 | 390174 | 1420 |
| 24 | 6.61 | 3.3 | 3.7 | 11 | 228557 | 5.36 | 307148 | 1014 |
| 19 | 7.22 | 3.4 | 3.9 | 11 | 231867 | 5.37 | 315130 | 1084 |
| 13 | 7.30 | 3.2 | 3.7 | 13 | 190267 | 5.28 | 238726 | 773 |
| 10 | 7.78 | 3.4 | 3.7 | 10 | 215323 | 5.33 | 308217 | 1036 |
| 33 | 8.08 | 3.3 | 3.9 | 15 | 212210 | 5.33 | 282131 | 937 |
| 54 | 9.00 | 3.4 | 3.8 | 10 | 311380 | 5.49 | 463398 | 1589 |
| 61 | 9.10 | 3.5 | 3.7 | 6 | 209390 | 5.32 | 276251 | 959 |
| 78 | 9.38 | 3.3 | 3.6 | 11 | 171547 | 5.23 | 207478 | 678 |
| 12 | 9.54 | 3.4 | 3.7 | 9 | 235483 | 5.37 | 317350 | 1066 |
| 32 | 9.56 | 3.4 | 3.8 | 9 | 211323 | 5.32 | 270950 | 929 |
| 77 | 10.04 | 3.3 | 3.8 | 12 | 158567 | 5.20 | 190204 | 633 |
| 71 | 10.60 | 3.4 | 3.7 | 8 | 195100 | 5.29 | 251898 | 864 |
| 46 | 10.68 | 3.6 | 3.8 | 8 | 274080 | 5.44 | 395278 | 1403 |

| 63 | 11.89 | 3.5 | 3.7 | 5 | 209357 | 5.32 | 276522 | 965 |
| 43 | 12.52 | 3.4 | 3.8 | 13 | 188577 | 5.28 | 272842 | 914 |
| 36 | 12.78 | 3.4 | 3.8 | 12 | 297543 | 5.47 | 448122 | 1515 |
| 39 | 18.05 | 3.4 | 3.8 | 13 | 198523 | 5.30 | 250829 | 840 |
| 23 | 18.57 | 3.4 | 3.9 | 13 | 180557 | 5.26 | 224803 | 755 |
| 52 | 19.33 | 3.4 | 3.9 | 12 | 250090 | 5.40 | 340323 | 1164 |
| 18 | 19.37 | 3.5 | 3.8 | 8 | 243267 | 5.39 | 333508 | 1164 |
| 25 | 20.93 | 3.6 | 3.9 | 6 | 242890 | 5.39 | 340841 | 1227 |
| 5 | 21.00 | 3.5 | 3.8 | 9 | 162343 | 5.21 | 198809 | 696 |
| 49 | 22.51 | 3.4 | 3.8 | 11 | 172110 | 5.24 | 213702 | 729 |
| 72 | 23.03 | 3.3 | 3.8 | 12 | 235180 | 5.37 | 317172 | 1047 |
| 53 | 23.20 | 3.4 | 3.9 | 14 | 186333 | 5.27 | 235507 | 794 |
| 73 | 26.89 | 3.4 | 3.7 | 9 | 202159 | 5.31 | 254249 | 864 |
| 1 | 27.30 | 3.6 | 3.8 | 6 | 277467 | 5.44 | 421277 | 1517 |
| 67 | 28.00 | 3.5 | 3.8 | 10 | 289547 | 5.46 | 422474 | 1479 |
| 27 | 32.68 | 3.5 | 3.8 | 8 | 265444 | 5.42 | 373680 | 1315 |
| 65 | 33.61 | 3.5 | 3.8 | 8 | 292112 | 5.47 | 432698 | 1514 |
| 57 | 51.99 | 3.4 | 3.8 | 11 | 222495 | 5.35 | 297515 | 1012 |
| 75 | 56.90 | 3.5 | 3.7 | 4 | 243707 | 5.39 | 339470 | 1202 |
| 22 | 61.19 | 3.4 | 3.7 | 9 | 205475 | 5.31 | 264775 | 900 |
| 2 | 80.70 | 3.5 | 3.7 | 7 | 241345 | 5.38 | 332898 | 1152 |
| 40 | 100.10 | 3.4 | 3.7 | 7 | 226634 | 5.36 | 303282 | 1031 |
| 4 | 106.10 | 3.5 | 3.7 | 5 | 269580 | 5.43 | 391587 | 1371 |
| 56 | 192.00 | 3.5 | | | 285979 | 5.46 | 429286 | 1515 |
| 21 | 208.70 | 3.5 | | | 211961 | 5.33 | 280271 | 989 |
| 66 | 246.60 | 3.6 | | | 296158 | 5.47 | 434452 | 1547 |
| 64 | 316.10 | 3.5 | | | 300380 | 5.48 | 450654 | 1559 |
| 55 | 2167.40 | 3.4 | | | 212867 | 5.33 | 280957 | 947 |

Fig. 1. Bulk and grain density as a function of the meteorite mass. Smaller samples show higher density scatter.

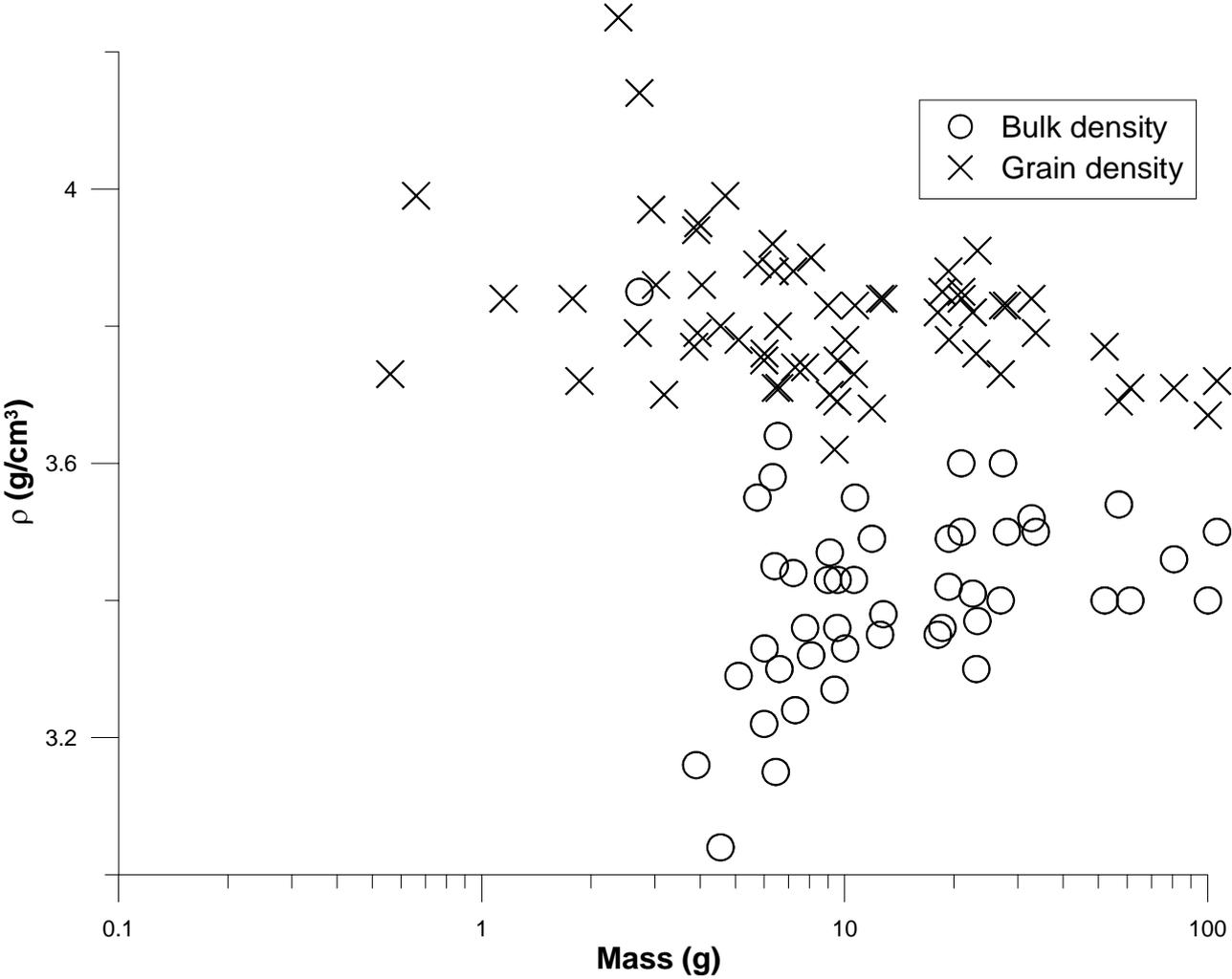

Fig. 2. Porosity as a function of the meteorite mass. Smaller samples show higher porosity scatter and larger samples tend to be less porous.

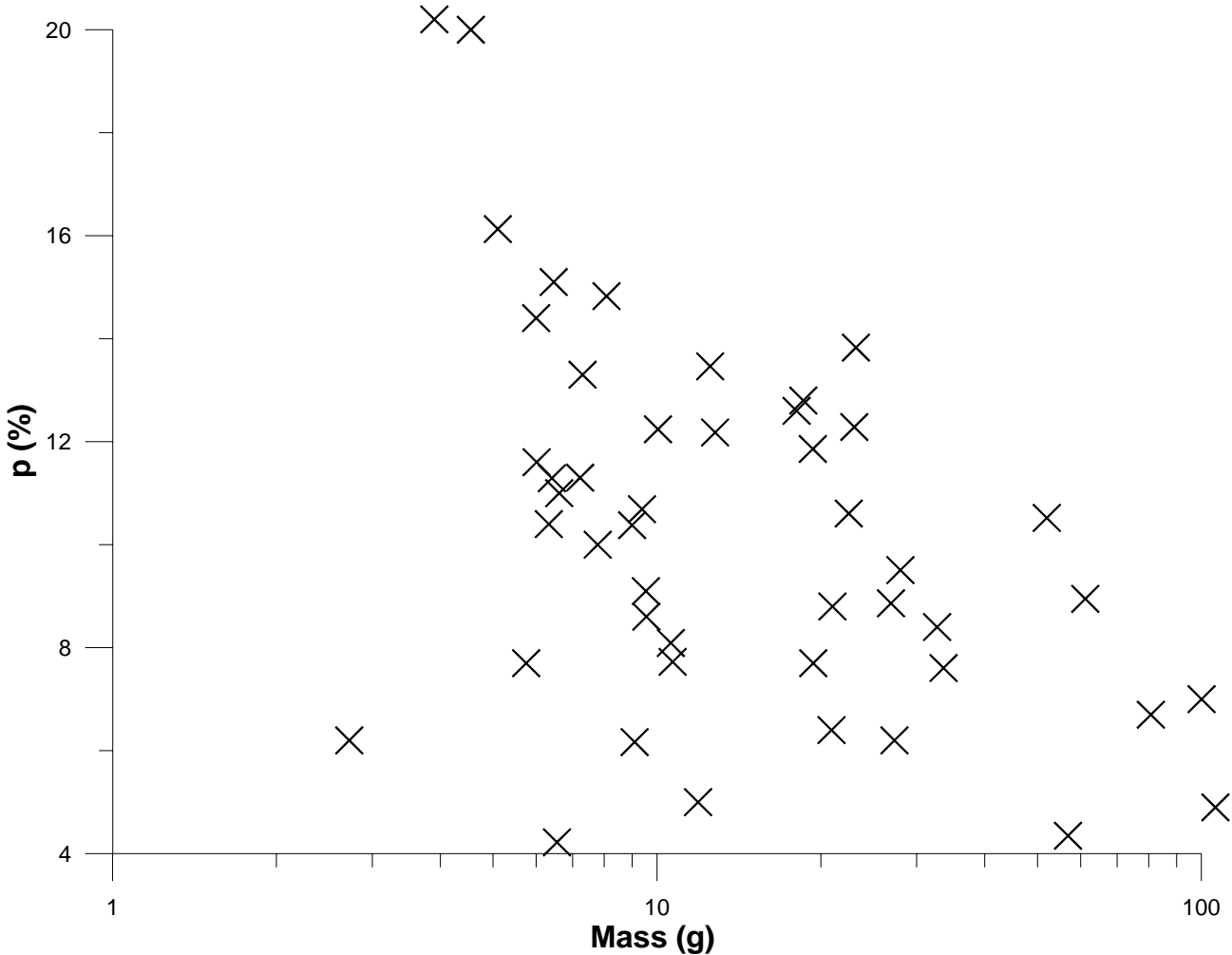

Fig. 3. Magnetic susceptibility as a function of the meteorite mass.

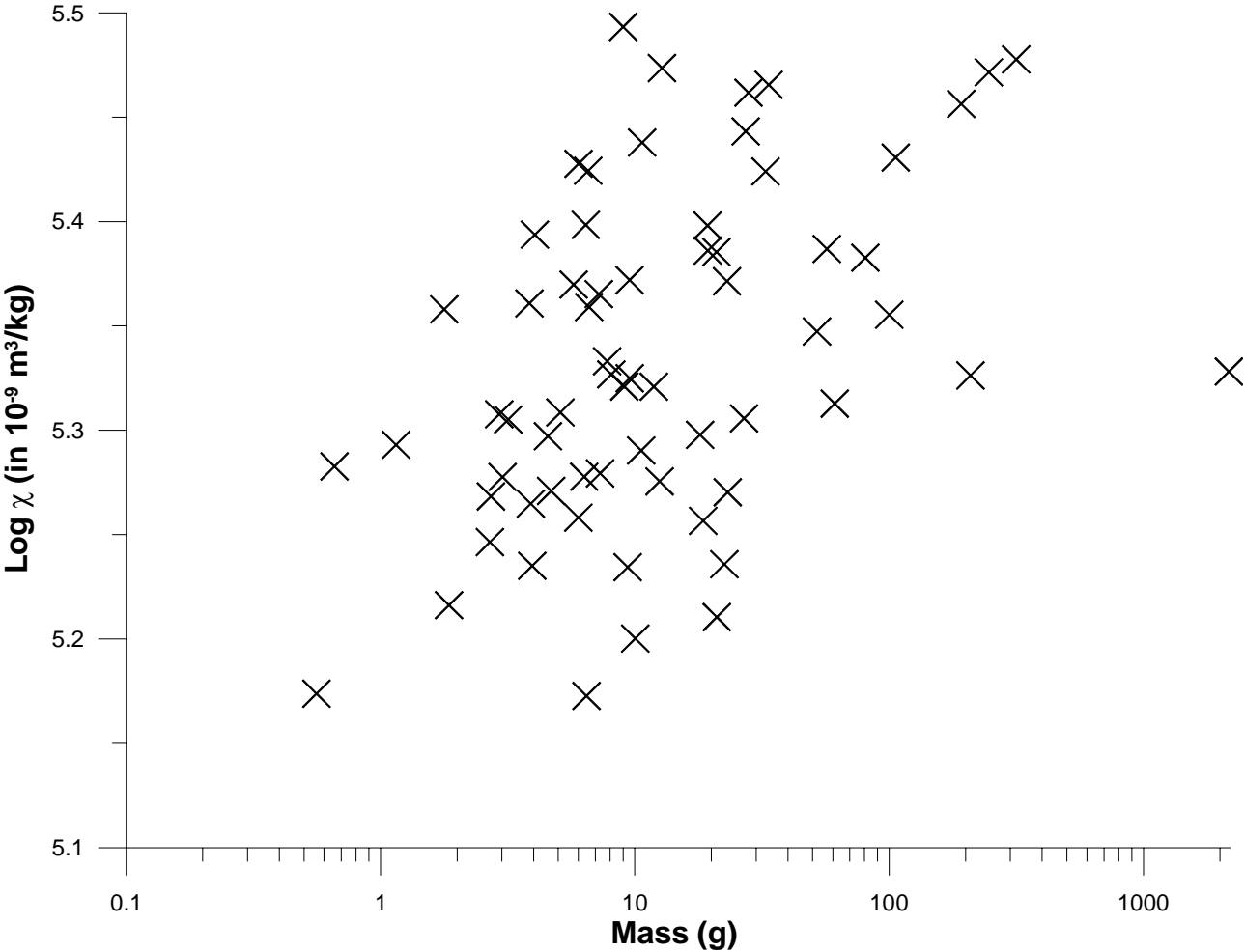

Fig. 4. Magnetic susceptibility as a function of the grain density.

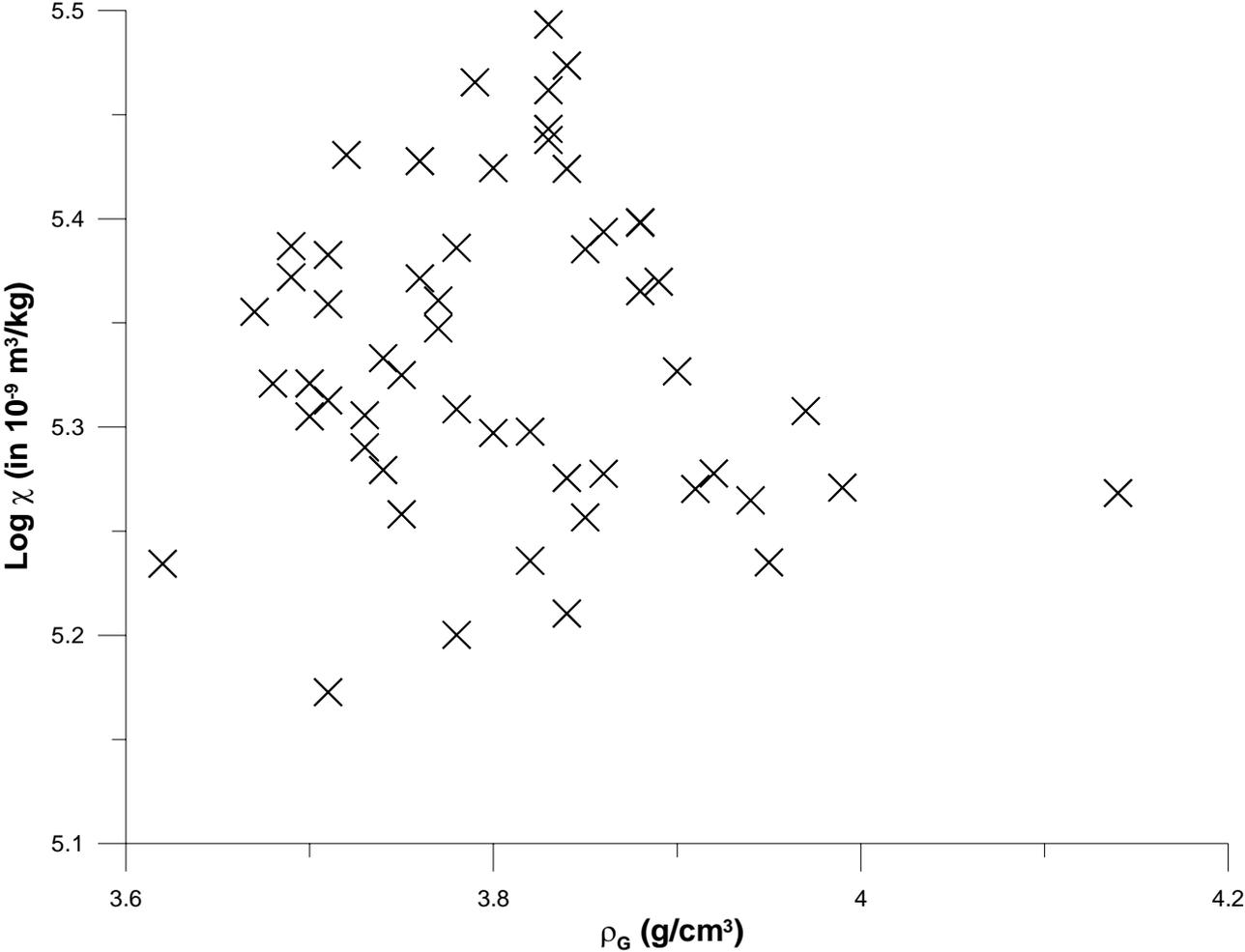